\documentclass[secnumarabic, pra]{revtex4}
\usepackage{amsmath}
\usepackage{bm}

\newcommand{\vA}{\mathbf{A}}
\newcommand{\vE}{\mathbf{E}}
\newcommand{\vnabla}{\bm{\nabla}}
\newcommand{\vx}{\mathbf{x}}
\newcommand{\vv}{\mathbf{v}}
\newcommand{\vs}{\mathbf{s}}

\begin{document}

\title{Energy conservation laws in classical electrodynamics}
\author{Valery P. Dmitriyev}
\affiliation{Lomonosov University\\
Box 160, Moscow, 117574, Russia}
\date{26 May 2004}

\begin{abstract}
There are three electromagnetic integrals of motion that can be
interpreted as the energy. These are the background energy, the
elastic energy and the integral in the torsion field commonly
referred to as the energy of the electromagnetic field. The
integral in the torsion field gains the meaning of the energy
insomuch as it is concerned with the mechanical energy of a
charged particle.
\end{abstract}

\maketitle

\section{Introduction. The electromagnetic field}

We will consider equations for electromagnetic potentials
\begin{eqnarray}
\frac {1}{c}\frac{\partial \vA}{\partial t} + \vE + \vnabla
\varphi = 0 , \label{1}\\
\vnabla \cdot \vE = 4 \pi q \delta (\vx - \vx ') ,\label{2}\\
\frac{\partial \vE}{\partial t} - c\,\vnabla \times (\vnabla \times
\vA) + 4 \pi q \vv \delta (\vx - \vx ') = 0 .\label{3}
\end{eqnarray}
The consideration will be restricted to the Coulomb gauge
\begin{equation}
\vnabla \cdot \vA = 0 . \label{4}
\end{equation}
We will find three integrals of equations (\ref{1}) $-$ (\ref{4})
that can be interpreted as the energy. This enables us to
elucidate the concept of the energy of the electromagnetic field.
Summation over recurrent index is implied throughout.

\section{The background energy}

Following \cite{Troshkin} we express the vector field $\vE$ via
some tensor field $\eta_{ik}$:
\begin{equation}
 E_i = \kappa \frac{\partial\eta_{ik}}{\partial x_k} , \label{5}
\end{equation}
where $\kappa$ is an arbitrary constant. Then (\ref{3}) can be obtained
convolving equation
\begin{equation}
\kappa \frac{\partial\eta_{ik}}{\partial t} + c \left(
\frac{\partial A_i}{\partial x_k} + \frac{\partial A_k}{\partial
x_i}\right) - q v_i \frac{\partial}{\partial x_k}\frac{1}{|\vx -
\vx'|} = 0 . \label{6}
\end{equation}
In derivation of (\ref{3})
from (\ref{6}) we used  (\ref{5}), (\ref{4}) and following
relations
\begin{eqnarray}
 \vnabla (\vnabla\cdot) = \vnabla^2 + \vnabla\times(\vnabla\times) , \label{7}\\
\vnabla^2\frac{1}{|\vx - \vx'|} = -\,4\pi \delta (\vx-\vx') .
\label{8}
\end{eqnarray}
Taking in (\ref{6}) $i=k$ and summing over the repeated index we get with the
account of (\ref{4})
\begin{equation}
\kappa \frac{\partial\eta_{kk}}{\partial t} - q v_k
\frac{\partial}{\partial x_k}  \frac{1}{|\vx - \vx'|} = 0 .
\label{9}
\end{equation}
Integrating (\ref{9}) all over the space
\begin{equation}
\frac{\partial}{\partial t}\int\eta_{kk}d^3x = 0 . \label{10}
\end{equation}
The quantity
\begin{equation}
\frac{1}{2}\eta_{kk} \label{11} \end{equation}
is interpreted in a mechanical model \cite{Troshkin} as the density of the background energy of a
substratum.

\section{The elastic energy}

Let us define the displacement field  $\vs$ by \vspace{-10pt}
\begin{equation}
\vA = \kappa c \frac{\partial \vs}{\partial t} . \label{12}
\end{equation}
Consider the case $\vv = 0$. Substituting (\ref{12}) into
(\ref{3}) and integrating it over time
\begin{equation}
\vE = \kappa c^2 \vnabla\times\vnabla\times\vs + \mathbf{h}(\vx) .
\label{13}
\end{equation}
Substituting (\ref{12}) and (\ref{13}) into (\ref{1}) we have
  by virtue of (\ref{2}) and (\ref{4})
\begin{eqnarray}
\frac{\partial^2\vs}{\partial t^2} +
c^2\vnabla\times\vnabla\times\vs = 0 , \label{14}\\
\mathbf{h} + \vnabla\varphi = 0 . \label{15}
\end{eqnarray}
Multiplying (\ref{14}) by $\partial \vs/\partial t$, integrating
over the space and taking the second integral by parts we get
\begin{equation}
\frac{\partial}{\partial t}\frac{1}{2}\int [(\frac{\partial
\vs}{\partial t})^2 + (c\vnabla\times\vs)^2]d^3x = 0 .
\label{16}
\end{equation}
This integral of motion is interpreted in the mechanical analogy
\cite{Dmitr} as the elastic energy of a substratum.

\section{Conservation in the torsion field}

Taking the curl of (\ref{14})
\begin{equation}
\frac{\partial^2(\vnabla\times\vs)}{\partial t^2} +
c^2\vnabla\times(\vnabla\times\vnabla\times\vs) = 0 . \label{17}
\end{equation}
Multiplying (\ref{17}) by $\partial (\vnabla\times\vs)/\partial
t$, integrating over the space and taking the second integral by
parts we get
\begin{equation}
\frac{\partial}{\partial t}\frac{1}{2}\int
[(\vnabla\times\frac{\partial \vs}{\partial t})^2 +
(c\vnabla\times\vnabla\times\vs)^2]d^3x = 0 . \label{18}
\end{equation}
Substituting (\ref{14}) and then (\ref{12}) into expression (\ref{18})
we convert it into the electromagnetic form
\begin{equation}
\frac{\partial}{\partial t}\frac{1}{2}\int [(\vnabla\times\vA)^2 +
(\frac {1}{c}\frac{\partial \vA}{\partial t})^2]d^3x = 0 .
\label{19}
\end{equation}

\section{The electromagnetic energy}

We will consider two charged particles. Forms (\ref{2}) and
(\ref{3}) are specified by
\begin{eqnarray}
\vnabla \cdot \vE = 4 \pi q_1 \delta (\vx - \vx^{(1)})+ 4 \pi q_2
\delta (\vx - \vx^{(2)}),\label{20}\\
\frac{\partial \vE}{\partial t} - c\,\vnabla \times (\vnabla \times
\vA) + 4 \pi q_1 \vv^{(1)} \delta (\vx - \vx^{(1)})+ 4 \pi q_2
\vv^{(2)} \delta (\vx - \vx^{(2)}) = 0 .\label{21}
\end{eqnarray}
The motion of the particles can be described by equations
\begin{eqnarray}
\vv^{(1)} = \frac{d\vx^{(1)}}{d t}, \qquad
m_1\frac{d\vv^{(1)}}{dt}
= \,q_1 \vE(\vx^{(1)}) + q_1\frac{\vv^{(1)}}{c}\times\vnabla\times\vA(\vx^{(1)}), \label{22}\\
\vv^{(2)} = \frac{d\vx^{(2)}}{d t}, \qquad
m_2\frac{d\vv^{(2)}}{dt} = \,q_2 \vE(\vx^{(2)}) +
q_2\frac{\vv^{(2)}}{c}\times\vnabla\times\vA(\vx^{(2)}).
\label{23}
\end{eqnarray}
Equations (\ref{22}) and (\ref{23}) close up the set of Maxwell's
equations (\ref{1}), (\ref{20}), (\ref{21}) and (\ref{4}).

Let us derive an integral of motion that is concerned with the
mechanical energy of the particles. Multiply (\ref{21}) by $\vE$,
then substitute (\ref{1}) into the second term. Integrate all over
the space and take the second integral by parts. This gives
\begin{equation}
\frac{\partial }{\partial t}\frac{1}{8\pi}\int[\vE^2 + (\vnabla
\times \vA)^2]d^3x + q_1 \vv^{(1)}\cdot \vE(\vx^{(1)}) + q_2
\vv^{(2)}\cdot \vE(\vx^{(2)}) = 0 .\label{24}
\end{equation}
Substitute (\ref{22}) and (\ref{23}) into (\ref{24}). Also for our
convenience we use (\ref{1}) in the first term of the expression
under the integral. Thus we get
\begin{equation}
\frac{\partial }{\partial t}\left\{\frac{1}{8\pi}\int[(\frac
{1}{c}\frac{\partial \vA}{\partial t} + \vnabla\varphi)^2 +
(\vnabla \times \vA)^2]d^3x +
\frac{1}{2}m_1\vv^{(1)}\cdot\vv^{(1)} +
\frac{1}{2}m_2\vv^{(2)}\cdot\vv^{(2)}\right\} = 0 .\label{25}
\end{equation}
The first term in (\ref{25}) is commonly interpreted as the energy
of the electromagnetic field. However, comparing (\ref{25}) with
(\ref{19}) we see that (\ref{25}) generalizes the integral of
motion in the torsion field of the displacement.

\section{Conclusion}

The electromagnetic unteraction has no relation to the background energy
nor to the elastic energy. It is concerned with a conservation law
 in the torsion field of a substratum.

\end{document}